\documentclass[aps,prd,onecolumn,groupedaddress,showpacs,nofootinbib,amssymb]{revtex4}
\usepackage{amsmath}
\usepackage{amssymb}
\usepackage{amsfonts}
\usepackage{graphicx,bm}
\usepackage{color,amsxtra}
\usepackage{epsf}
\usepackage{enumerate}
\usepackage{hhline}
\usepackage{array}
\usepackage{tabularx}
\newcommand{\be}{\begin{equation}}
\newcommand{\ee}{\end{equation}}
\newcommand{\bea}{\begin{eqnarray}}
\newcommand{\eea}{\end{eqnarray}}
\newcommand{\beaa}{\begin{eqnarray*}}
\newcommand{\eeaa}{\end{eqnarray*}}

\newcommand{\nn}{\nonumber \\}
\newcommand{\e}{\mathrm{e}}

\def\be{\begin{equation}}
\def\ee{\end{equation}}
\def\bea{\begin{eqnarray}}
\def\eea{\end{eqnarray}}

\def\nn{\nonumber \\}
\def\e{\mathrm{e}}

\allowdisplaybreaks[4]

\begin{document}

\tolerance=5000

\title{Localization of Vector Field on Dynamical Domain Wall}

\author{Masafumi Higuchi$^{1, }$\footnote{
E-mail address: mhiguchi@th.phys.nagoya-u.ac.jp}, 
Shin'ichi Nojiri$^{1, 2,}$\footnote{E-mail address:
nojiri@phys.nagoya-u.ac.jp}
}

\affiliation{
$^1$ Department of Physics, Nagoya University, Nagoya
464-8602, Japan \\
$^2$ Kobayashi-Maskawa Institute for the Origin of Particles and
the Universe, Nagoya University, Nagoya 464-8602, Japan}

\begin{abstract}

In the previous works (arXiv:1202.5375 and 1402.1346), the dynamical domain 
wall, where the four dimensional FRW universe is embedded in the five 
dimensional space-time, has been realized by using two scalar fields. 
In this paper, we consider the localization of vector field in three formulations. 
The first formulation was investigated in the previous paper 
(arXiv:1510.01099) for the $U(1)$ gauge field.  
In the second formulation, we investigate the Dvali-Shifman mechanism 
(hep-th/9612128), where the non-abelian gauge field is confined in the bulk 
but the gauge symmetry is spontaneously broken on the domain wall. 
In the third formulation, we investigate the Kaluza-Klein modes coming from 
the five dimensional graviton. In the Randall-Sundrum model, the graviton was 
localized on the brane. We show that the $(5,\mu)$ components 
$\left(\mu=0,1,2,3\right)$ of the graviton are also localized on the domain 
wall and can be regarded as the vector field on the domain wall. 
There are, however, some corrections coming from the bulk extra dimension 
if the domain wall universe is expanding. 

\end{abstract}


\maketitle

\section{Introduction \label{I}}

There is a long history in the scenarios that our universe could be a brane 
or domain wall embedded in a higher dimensional space-time \cite{Akama:1982jy, Rubakov:1983bb}. 
After the discovery of the so-called $D$-brane solution in string 
theories \cite{Dai:1989ua, Polchinski:1996na}, the brane world scenarios 
\cite{Randall:1999ee, Randall:1999vf, Dvali:2000hr, Deffayet:2000uy, Deffayet:2001pu} or the domain wall senario  
\cite{Lukas:1998yy, Kaloper:1999sm, Chamblin:1999ya, DeWolfe:1999cp, Gremm:1999pj, Csaki:2000fc,Gremm:2000dj, Kehagias:2000au, Kobayashi:2001jd, Slatyer:2006un, George:2008vu} 
have been well studied. 
In the studies, the models of the inflationary brane using the trace anomaly 
have been proposed \cite{Nojiri:2000eb,Hawking:2000bb,Nojiri:2000gb}. 
The brane can be regarded with a limit that the thickness of the domain wall 
vanishes. 
Recently a model where the general FRW universe is embedded in the five 
dimensional space-time with an arbitrary warp factor by using two scalar 
fields \cite{Toyozato:2012zh, Higuchi:2014bya} \footnote{
This formulation is an extension of the formalism of the reconstruction of 
the domain wall \cite{Nojiri:2010wj}. 
Before the work, a formulation where only 
the warp factor of the domain wall is arbitrary has been proposed in 
\cite{DeWolfe:1999cp}, it has been proposed. }

In this paper, we investigate the localization of the vector field in the model 
\cite{Toyozato:2012zh, Higuchi:2014bya} by using three formulations. 
The localization of the graviton has been shown in \cite{Higuchi:2014bya} 
and the localizations of the spinor field and vector field have been also 
investigated and shown in \cite{Toyozato:2015xfa}. 
The first formulation of the localization of the vector field in this paper 
is just the review of the work in \cite{Toyozato:2015xfa} about the 
$U(1)$ gauge field by using the action of the five dimensional vector field. 
The second formulation is an extension of the Dvali-Shifman mechanism 
in \cite{Dvali:1996xe}, where 
the non-abelian gauge field is confined in the four dimensional bulk  
but the gauge symmetry is spontaneously broken on the three dimensional 
domain wall. 
An extension of the work in \cite{Dvali:1996xe} on the static four dimensional 
domain wall has been investigated in \cite{Davies:2007xr} and in this paper, 
we further extend the mechanism to the dynamical domain wall model. 
As the third formulation, we investigate the Kaluza-Klein modes coming from 
the five dimensional graviton. 
In the third formulation, we consider the vector field coming from 
the Kaluza-Klein reduction. 
In the second Randall-Sundrum model \cite{Randall:1999vf}, the graviton was 
localized on the brane. The localized graviton can be regarded as a zero mode 
of the five dimensional graviton. 
We show that the $(5,\mu)$ components $\left(\mu=0,1,2,3\right)$ of the 
graviton are also localized and can be regarded as the vector field on the four 
dimensional domain wall. 
We show that, however, there appear some corrections coming from the bulk 
extra dimension if the domain wall is dynamical. 

In the next section, we briefly review on the formulation of the dynamical 
domain wall based on \cite{Toyozato:2012zh, Higuchi:2014bya}. 
In section \ref{III}, we also review on the localization of the vector field 
in \cite{Toyozato:2015xfa}. 
In section \ref{IV}, we extend the formulation in \cite{Dvali:1996xe} and 
\cite{Davies:2007xr} to the four dimensional dynamical domain wall model.
In section \ref{V}, we consider the Kaluza-Klein vector field coming from the 
five dimensional graviton. 
The last section \ref{VI} is devoted to the summary of the obtained results. 

\section{Domain wall model with two scalar fields \label{II}}

In \cite{Toyozato:2012zh, Higuchi:2014bya}, the formulation of the dynamical 
domain wall model have been proposed by using two scalar fields. 
The formulation could be regarded as an extension of the formulation 
in \cite{Bamba:2011nm} \footnote{A similar procedure was also invented 
for the reconstruction of the FRW universe by single scalar model 
\cite{Capozziello:2005tf}.}. 

The metric of the five dimensional space-tme embedded a general 
spacially flat FRW universe with an arbitrary warp factor is given by 
\be
\label{metricR}
ds^2 = dw^2 + L^2 \e^{u \left( w,t \right)} ds_\mathrm{FRW}^2 \, . 
\ee
Here $ds_\mathrm{FRW}^2$ is the metric of the FRW universe, 
\be
\label{FRWmetric0}
ds_\mathrm{FRW}^2 = - dt^2 + a \left( t \right)^2 \left\{ dr^2
+ r^2 d \theta^2 +r^2 \sin^2 \theta d \phi^2  \right\}\, .
\ee
In \cite{Toyozato:2012zh, Higuchi:2014bya}, the following action with 
two scalar fields $\phi$ and $\chi$ were considered, 
\be
\label{pc1}
S_{\phi\chi} = \int d^5 x \sqrt{-g} \left\{ \frac{R}{2\kappa^2} 
 - \frac{1}{2} A (\phi,\chi) \partial_M \phi \partial^M \phi 
 - B (\phi,\chi) \partial_M \phi \partial^M \chi 
 - \frac{1}{2} C (\phi,\chi) \partial_M \chi \partial^M \chi - V (\phi,\chi)\right\}\, .
\ee
We can construct a model to realize the arbitrary 
metric (\ref{metricR}) by using the model (\ref{pc1}).
The energy-momentum tensor for the scalar fields $\phi$ and $\chi$ 
in the model (\ref{pc1}) are given by
\begin{align}
\label{pc2}
T^{\phi\chi}_{MN} =& g_{MN} \left\{ 
 - \frac{1}{2} A (\phi,\chi) \partial_L \phi \partial^L \phi 
 - B (\phi,\chi) \partial_L \phi \partial^L \chi 
 - \frac{1}{2} C (\phi,\chi) \partial_L \chi \partial^L \chi - V (\phi,\chi)\right\} \nn
& +  A (\phi,\chi) \partial_M \phi \partial_N \phi 
+ B (\phi,\chi) \left( \partial_M \phi \partial_N \chi 
+ \partial_N \phi \partial_M \chi \right) 
+ C (\phi,\chi) \partial_M \chi \partial_N \chi \, .
\end{align}
The variations of $\phi$ and $\chi$ give the following field equations, 
\begin{align}
\label{pc3}
0 =& \frac{1}{2} A_\phi \partial_M \phi \partial^M \phi 
+ A \nabla^M \partial_M \phi + A_\chi \partial_M \phi \partial^M \chi 
+ \left( B_\chi - \frac{1}{2} C_\phi \right)\partial_M \chi \partial^M \chi  
+ B \nabla^M \partial_M \chi - V_\phi \, ,\\
\label{pc4}
0 =& \left( - \frac{1}{2} A_\chi + B_\phi \right) \partial_M \phi \partial^M \phi 
+ B \nabla^M \partial_M \phi 
+ \frac{1}{2} C_\chi \partial_M \chi \partial^M \chi 
+ C \nabla^M \partial_M \chi + C_\phi \partial_M \phi \partial^M \chi 
 - V_\chi\, .
\end{align}
Here $A_\phi=\partial A(\phi,\chi)/\partial \phi$, etc. 
By choosing $\phi=t$ and $\chi=w$, we obtain 
\be
\label{pc4b}
T_0^{\ 0} = - \frac{\e^{-2u \left( w,t \right)} }{2L^2 } A - \frac{1}{2} C - V\, ,
\quad 
T_i^{\ j} = \delta_i^{\ j} \left( \frac{\e^{-2u \left( w,t \right)} }{2L^2 } A
 - \frac{1}{2} C - V \right)\, ,\quad 
T_5^{\ 5} = \frac{\e^{-2u \left( w,t \right)} }{2L^2 } A + \frac{1}{2} C - V\, ,
\quad 
T_0^{\ 5} = B \, ,
\ee
By using the Einstein equation and the equations in (\ref{pc4b}), 
we find $A$, $B$, $C$, and $V$ can be expressed as follows,  
\begin{align}
\label{pc7}
A =& \frac{L^2 \e^{u \left( w,t \right)}}{\kappa^2 } 
\left( G_1^{\ 1} - G_0^{\ 0} \right) 
=  \frac{L^2 \e^{u \left( w,t \right)}}{\kappa^2 } 
\left( G_2^{\ 2} - G_0^{\ 0} \right) 
=  \frac{L^2 \e^{u \left( w,t \right)}}{\kappa^2 } 
\left( G_3^{\ 3} - G_0^{\ 0} \right) \nn
=& \frac{1}{\kappa^2} \left( - \ddot u - 2 \dot H 
+ \frac{\left( \dot u \right)^2}{2} + \dot u H \right) \, , \nn
B = & \frac{1}{\kappa^2}G_0^{\ 5} = - \frac{3u'}{2 \kappa^2 L^2 \e^u} 
\left( \dot u + 2 H \right) \, , \nn
C =& \frac{1}{\kappa^2} \left( G_5^{\ 5} - G_1^{\ 1} \right)
= \frac{1}{\kappa^2} \left( G_5^{\ 5} - G_2^{\ 2} \right) 
= \frac{1}{\kappa^2} \left( G_5^{\ 5} - G_3^{\ 3} \right) \nn
=& \frac{1}{\kappa^2} \left( - \frac{3}{2} u'' - \frac{1}{2\e^u} \left( \ddot u 
+ 2 \dot H + \left( \dot u \right)^2 +5\dot u H + 6 H^2 \right) \right)\, , \nn
V = & \frac{1}{\kappa^2} \left( G_0^{\ 0} + G_5^{\ 5} \right) \nn
=& \frac{1}{\kappa^2} \left( - \frac{3}{4} \left( u'' + 2 \left( u' \right)^2 \right) 
+ \frac{1}{4L^2 \e^u} \left( 3 \ddot u + 6 \dot H + 3 \left( \dot u \right)^2 
+ 15 \dot u + 18 H^2 \right) \right)\, .
\end{align}
Here $G_{\mu\nu}$ is the Einstein tensor. 
The explicit forms of $A(\phi,\chi)$, 
$B(\phi,\chi)$, $C(\phi,\chi)$, and $V(\phi,\chi)$ can be obtained 
by replacing $t$ and $w$ in the r.h.s. of Eqs.~(\ref{pc7}) by $\phi$ and $\chi$. 
The obtained expressions in the action (\ref{pc1}) gives a model 
which realize the metric (\ref{metricR}). 
Eqs.~(\ref{pc3}) and (\ref{pc4}) are satisfied automatically, which can be seen 
by using the Bianchi identity 
$\nabla^\nu \left( R_{\mu\nu} - \frac{1}{2} R g_{\mu\nu} \right) = 0$. 

\section{Localization of Vector Field \label{III}}

In this section, we review on the localization of the vector field 
by using the formulation in \cite{Toyozato:2015xfa}. 
We consider the following action of the five dimensional vector field, 
\be
\label{V1}
S_V = \int d^5 x \sqrt{-g} \left\{ - \frac{1}{4} F_{MN} F^{MN} - \frac{1}{2} m(\chi)^2 A_M A^M \right\}\, , \quad 
F_{MN} = \partial_M A_N - \partial_N A_M\, .
\ee
In the background (\ref{metricR}) with (\ref{FRWmetric0}), we assume that 
$\e^{u\left(  t, w \right)}$ is given by the product of the $t$-dependent part 
and $w$-dependent part, 
$\e^{u\left(  t, w \right)} = T \left( t \right) W \left( w \right)$, 
\be
\label{metricRB}
ds^2 = dw^2 + L^2 W(w) T(t) ds_\mathrm{FRW}^2 \, , \quad 
ds_\mathrm{FRW}^2 = - dt^2 + a(t)^2 \sum_{i=1}^3 \left( dx^i \right)^2 \, .
\ee
Under the assumption (\ref{metricRB}), the action (\ref{V1}) has the 
following form, 
\begin{align}
\label{V2}
S_V =& \int d^5 x \left\{ \frac{1}{2} L^2 W(w) T(t) a(t)^3 F_{50}^2 
 - \frac{1}{2} L^2 W(w) T(t) a(t) F_{5i}^2 
+ \frac{1}{2}a(t) F_{0i}^2 - \frac{1}{4} a(t)^{-1} F_{ij}^2 \right. \nn
& \left. - \frac{1}{2} m(\chi)^2 \left( L^4 W(w)^2 T(t)^2 a(t)^3 A_5^2 
 - L^2 W(w) T(t) a(t)^3 A_0^2 + L^2 W(w) T(t) a(t) A_i^2 \right) \right\}\, .
\end{align}
The variations of $A_5$, $A_0$, and $A_i$ give the following equations,
\begin{align}
\label{V3}
0 = & L^2 W(w) \partial_0 \left( T(t) a(t)^3 \left( \partial_5 A_0 - \partial_0 A_5 \right) \right) - L^2 W(w) T(t) a(t) \left(\partial_5 \partial_i A_i - \partial_i^2 A_5 \right) \nn
& - m(\chi)^2 L^4 W(w)^2 T(t)^2 a(t)^3 A_5 \, , \\
\label{V4}
0 = & - L^2 T(t) a(t)^3 \partial_5 \left( W(w) \left( \partial_5 A_0 
 - \partial_0 A_5 \right) \right) + a(t) \left( \partial_0 \partial_i A_i 
 - \partial_i^2 A_0 \right) + m(\chi)^2 L^2 W(w) T(t) a(t)^3 A_0\, , \\
\label{V5}
0 = & L^2 T(t) a(t) \partial_w \left( W(w) 
\left( \partial_5 A_i - \partial_i A_5 \right) \right) 
 - \partial_0 \left( a \left( \partial_0 A_i - \partial_i A_0 \right) \right) 
 - a(t)^{-1}  \left( \partial_i \partial_j A_j - \partial_j^2 A_i \right) \nn
& - m(\chi)^2 L^2 W(w) T(t) a(t) A_i \, .
\end{align}
If we assume 
\be
\label{V6} 
A_5 =0 \, , \quad 
A_\mu = X(w) C_\mu \left( x^\nu \right)\, , \quad 
\mu,\nu=0,1,2,3\, ,
\ee
and choose  
\be
\label{V7}
m\left( \chi=w \right)^2 = \frac{\left( W (w) X'(w) \right)'}{W(w) X(w)}\, ,
\ee
we rewrite Eqs.~(\ref{V3}), (\ref{V4}), and (\ref{V5}) as follows, 
\begin{align}
\label{V8}
0 =& \partial_5 X(w) \left\{ \partial_0 \left(T(t) a(t)^3 C_0 \right) 
 - T(t) a(t) \partial_i C_i \right\}\, , \\
\label{V9}
0 = & \partial_0 \partial_i C_i - \partial_i^2 C_0 \, , \\
\label{V10}
0 = &
\partial_0 \left( a(t) \left( \partial_0 C_i - \partial_i C_0 \right) \right) 
+ a(t)^{-1} \left( \partial_i \partial_j C_j - \partial_j^2 C_i \right) \, .
\end{align}
Eqs.~(\ref{V9}) and (\ref{V10}) are nothing but the field equations 
of the vector field in four dimensions. 
On the other hand, Eq.~(\ref{V8}) can be regarded as a gauge condition, 
which is a generalization of the Landau gauge, $\partial^\mu A_\mu = 0$.  

By choosing $X(w)$ decreases rapidly enough for large $\left|w \right|$, 
$A_\mu$ becomes normalizable. 
Then if we choose $m(\chi)$ as in (\ref{V7}), the vector field localizes 
on the domain wall. 

\section{Dvali-Shifman mechanism \label{IV}}

In \cite{Dvali:1996xe}, the non-abelian vector field on the three dimensional 
domain wall embedded in the four dimensional space-time was considered. 
In the bulk space-time, the vector field is confined but on the domain 
wall, the scalar field which generates the domain wall also change the 
potential of the Higgs field and there occurs the spontaneous breakdown 
of the gauge 
symmetry and massless $U(1)$ gauge field appears on the domain wall. 
An extension of the scenario was proposed in \cite{Davies:2007xr}, where  
the four dimensional domain wall in the five dimensional space-time was 
considered and by the mechanism similar to that in \cite{Dvali:1996xe}, 
the standard model could be realized on the domain wall. 
In this section, we consider a similar mechanism on the dynamical domain 
wall. 

We consider the following action for the $SU(2)$ gauge field, 
\begin{align}
\label{actiong}
S=&\int dx^5\sqrt{-g}\left[-\frac{1}{4}G^a_{MN}G^a_{MN}
 -\frac{1}{2}(D_M\eta^a)^2+\frac{1}{2}\lambda \left(\eta^2
+\kappa^2-v^2+v^2\tanh^2(m\chi) \right)^2\right] \, ,\nn
G^a_{MN}=&\partial_M A^a_N-\partial_N A^a_M+gf^{abc}A^b_M A^c_N \, .
\end{align}
Here $G^a_{MN}$ is the field strength of the $SU(2)$ field and we also include 
the scalar field $\eta^a$ which is the adjoint representation of $SU(2)$. 
The parameters $\kappa$ and $v$ have the dimension of mass and 
$\lambda$ is a dimensionless positive parameter. 
We assume $\kappa^2-v^2<0$ and also the metric in (\ref{metricRB}). 

In the limit of $|w|=|\chi| \to\infty$, because the potential for the scalar field 
$\eta^a$ is given by $\frac{1}{2}\lambda(\eta^2+\kappa^2)^2$, there does not 
occur the breakdown of the $SU(2)$ gauge symmetry if the 
gauge coupling is strong enough. 
On the other hand, on the brane, $w=\chi \sim0$, the potential becomes 
$\frac{1}{2}\lambda(\eta^2+\kappa^2-v^2)^2$ and because 
$\kappa^2-v^2<0$, $\eta^a$ has a vacuum expectation value 
\begin{align}
\eta^a=\delta_{3a}\eta_0(w)=\delta_{3a}k\cosh^{-1}(mw)\, ,
\label{190649_24Oct16}
\end{align}
and therefore $SU(2)$ guge symmetry is spontaneously broken. 
By substituting the expression of $\eta^a$ into (\ref{190649_24Oct16}) 
into the equation of the motion 
\begin{align}
 -\eta^{\prime\prime}_0+2\lambda
(\eta^2+\kappa^2-v^2+v^2\tanh^2(m\phi))\eta_0=0\, ,
\label{123725_22Oct16}
\end{align}
we obtain 
\be
\tanh^2(mw)\left(-2m^2-2\lambda k^2+2\lambda v^2\right)+m^2+2\lambda(k^2+\kappa^2-v^2)=0 \, ,
\ee
which tells
\be
k^2=v^2-2\kappa^2,\ m^2=2\lambda\kappa^2 \, .
\ee
Because the $\eta^a$ has a vacuum expectation value, 
\begin{align}
\begin{pmatrix}
\langle\eta^1\rangle\\
\langle\eta^2\rangle\\
\langle\eta^3\rangle
\end{pmatrix}=
\begin{pmatrix}
0\\
0\\
\left[(1-\tanh^2(mw))v^2-\kappa^2\right]^{1/2} 
\end{pmatrix} \, ,
\label{160158_25Oct16}
\end{align}
the gauge field obtains a mass, 
\begin{align}
\Delta\mathcal{L}=&
 -\frac{1}{2}g^2f^{abc}f^{ab^\prime c^\prime}
g^{MN}A^b_M A^{b^\prime}_N\eta^c\eta^{c^\prime}\, .
\end{align}
Then by using (\ref{160158_25Oct16}), we find
\begin{align}
\Delta\mathcal{L}=&-\frac{1}{2}g^2g^{MN}(A^1_M A^1_N+A^2_M A^2_N)
\langle\eta^3\rangle^2 \nn
=&-\frac{1}{2}g^2g^{MN}(A^1_M A^1_N+A^2_M A^2_N)
\left[(1-\tanh^2(mw))v^2-\kappa^2\right]\nn 
=&-\frac{1}{2}\mu(w)^2(1-\delta_{3a})g^{MN}A^a_M A^a_N \, .
\end{align}
By substituting (\ref{metricRB}) into the action (\ref{actiong}), 
we obtain 
\begin{align}
\label{actiong2}
S=&\int d^5x\left[\frac{1}{2}a(t)(G^a_{0i})^2+\frac{1}{2}
L^2 W(w)T(t)a(t)^3(G^a_{05})^2-\frac{1}{4}a(t)^{-1}(G^a_{ij})^2 \right. \nn 
&-\frac{1}{2}L^2 W(w)T(t)a(t)(G^a_{i5})^2  
 -\frac{1}{2}\mu(w)^2 L^2W(w)T(t)a(t) \left[-a(t)^2\left((A^1_0)^2
+(A^2_0)^2\right) \right. \nn 
& \left. \left. +\left((A^1_i)^2+(A^2_i)^2\right)
+L^2 W(w)T(t)a(t)^2\left((A^1_5)^2 +(A^2_5)^2\right)\right] \right]\, .
\end{align}
Then the equations for the gauge fields are given by 
\begin{align}
\label{gf1}
0=&T(t)a(t)^3\partial_5 \left[L^2 W(w)(\partial_0A^a_5-\partial_5A^a_0)\right]
+a(t)(\partial_0\partial_iA^a_i-\partial^2_iA^a_0) \nn 
& +(1-\delta_{a3})\mu(w)^2 L^2 W(w)T(t)a^3(t)A^a_0 
+gf^{abc}\bigl[a(t)(A^b_0\partial_iA^c_i-A^c_i\partial_0A^b_i
+2A^c_i\partial_iA^b_0)\nn 
&+L^2 W(w)T(t)a(t)^3(A^b_0\partial_5A^c_5-A^c_5\partial_0A^b_5+2A^c_5
\partial_5A^b_0)+W^\prime(w)
T(t)a(t)^3A^b_0A^c_5\bigr]\nn 
&+g^2f^{abc}f^{cde}\left[a(t)A^b_iA^d_0A^e_i
+L^2 W(w)T(t)a(t)^3A^b_5A^d_0A^e_5\right] \, ,\\
\label{gf2}
0=&T(t)a(t)\partial_5\left[L^2 W(w)(\partial_5A^a_i-\partial_iA^a_5)
\right]-\partial_0\left[a(t)(\partial_0A^a_i-\partial_iA^a_0)
\right]-a(t)^{-1}(\partial_i\partial_jA^a_j-\partial^2_jA^a_i) \nn 
&-(1-\delta_{a3})\mu(w)^2 L^2 W(w)T(t)a(t)A^a_i 
 -gf^{abc}\bigl[a(t)(A^c_i\partial_0A^b_0
+A^c_0\partial_iA^b_0+2A^b_0\partial_0A^c_i)+\dot{a}(t)A^b_0A^c_i\nn 
&-a(t)^{-1}(A^c_i\partial_jA^b_j+A^c_j\partial_iA^b_j+2A^b_j\partial_jA^c_i) \nn 
&+ L^2 W(w)T(t)a(t)(A^b_i\partial_5A^c_5-A^c_5\partial_iA^b_5
+2A^c_5\partial_5A^b_i)+ L^2 W(w)^\prime T(t)a(t)A^b_iA^c_5\bigr]\nn 
&-g^2f^{abc}f^{cde}(a(t)A^b_0A^d_0-a(t)^{-1}A^b_jA^d_j- L^2 W(w)
T(t)a(t)A^b_5A^d_5)A^e_i \, ,\\
\label{gf3}
0=&L^2 W(w)T(t)a(t)\left[\partial^2_iA^a_5-\partial_5\partial_iA^a_i\right]
 -L^2 W(w)\partial_0\left[T(t)a(t)^3(\partial_0A^a_5-\partial_5A^a_0)\right]\nn 
&-(1-\delta_{a3})\mu(w)^2 L^2 W(w)^2T(t)^2a(t)^3A^a_5 \nn 
&-gf^{abc}\bigl[L^2 W(w)T(t)a(t)^3(A^c_5\partial_0A^b_0
+A^c_0\partial_5A^b_0+A^b_0\partial_0A^c_5)
+L^2 W(w)\partial_0(T(t)a(t)^3)A^b_0A^c_5\nn 
&-W(w)T(t)a(t)(A^c_5\partial_iA^b_i
+A^c_i\partial_5A^b_i+A^b_i\partial_iA^c_5)\bigr] \nn 
&-g^2f^{abc}f^{cde}\left[L^2 W(w)T(t)a(t)^3A^b_0A^d_0
+L^2 W(w)T(t)a(t)A^b_iA^d_i\right]A^e_5 \, .
\end{align}
For the massless vector field $A^3_M$, by choosing 
\begin{align}
A^3_5=0, \quad A^3_\mu=X(w)C_\mu(x^\nu)\, , \quad\mu=0,1,2,3\, ,
\end{align}
and 
\begin{align}
 (W(w)X^\prime(w))^\prime=0 \, ,
\end{align}
in the order of $\mathcal{O}\left( g^0 \right)$, 
Eqs.~(\ref{gf1}), (\ref{gf2}), and (\ref{gf3}) reduce to the equations for the 
vector field and the gauge fixing condition, 
\begin{align}
0=&a(t)X(w)(\partial_0\partial_iC_i-\partial^2_iC_0) \, ,\\
0=&\partial_0\left[a(t)(\partial_0C_i-\partial_iC_0)\right]
+a(t)^{-1}(\partial_i\partial_jC_j-\partial^2_jC_i) \, ,\\
0=&X^\prime(w)\left[-T(t)a(t)\partial_iC_i+\partial_0(T(t)a(t)^3C_0)\right] \, .
\end{align}
Therefore the massless gauge field appears on the domain wall. 

In \cite{Davies:2007xr}, the confinement in the bulk space-time was assumed 
but in the dimensions higher than four, there could be a phase transition 
and the confinement could occur only in the strong coupling region. 
Then we may consider the scalar field, which also plays a role of the gauge 
coupling. 
The scalar field depends on the coordinate $w$ in the extra dimension and 
the gauge coupling can become strong and the confinement always occurs 
in the bulk space-time.

\section{Kaluza-Klein reduction \label{V}}

In the Randall-Sundrum model, the massless graviton in four dimensions 
appears as a zero mode, or normalized and localized mode, of 
the five dimensional graviton. 
On the other hand, in the standard Kaluza-Klein model, the vector field 
appears as the fluctuation $h_{5\mu}$ of the $(5,\mu)$ components of 
the metric $\left( \mu = 0,1,2,3 \right)$. 
Therefore if the $(5,\mu)$ components are also localized on the brane or the 
domain wall, the modes can be regarded as the vector field in four 
dimensions. 
In this section, we investigate the possiblity that the vector field appears 
due to the Kaluza-Klein reduction. 

We consider the fluctuation arround the background space-time, 
$g_{AB}=g^{(0)}_{AB}+h_{AB}$. 
Then we obtain the following expressions, 
\begin{align}
\sqrt{-g}=&\sqrt{-g^{(0)}}\left(1+\frac{1}{2}h^{A}_{A}
+\frac{1}{8}(h^A_A)^2-\frac{1}{4}h_{AB}h^{AB}\right) \, ,\\
R=&R^{(0)}-R^{(0)AB}h_{AB}+\nabla^{(0)A}\nabla^{(0)B}h_{AB}
 -{\nabla^{(0)}}^2h_A^{\ A}\nn 
&+\frac{3}{4}\left(\nabla^{(0)C}h_A^{\ A}\right)\nabla^{(0)B}h_{CB}
+\frac{3}{4}h^{AB}\nabla^{(0)}_A\nabla^{(0)}_Bh_D^{\ D}
 -\frac{1}{2}\nabla^{(0)A}h_{AE}\nabla^{(0)}_Bh^{BE} \nn 
&+\frac{1}{4}h^{AB}{\nabla^{(0)}}^2h_{AB}
 -\frac{1}{4}\left(\nabla^{(0)C}h_D^{\ D}\right)
\left(\nabla^{(0)}_Ch_A^{\ A}\right)+\frac{1}{2}R^{(0)}_{AB}h^{AD}h^B_{\ D}
+\frac{1}{2}R^{(0)DACB}h_{DC}h_{AB} \nn 
& +\nabla^{(0)A}\left(-\frac{1}{4}h_{FB}\nabla_Ah^{BF}-\frac{1}{2}h_{AF}\nabla^{(0)}_Dh^{DF}+\frac{1}{4}h_{AF}\nabla^{(0)F}h_D^{\ D}\right) \nn 
&+\frac{1}{2}{\nabla^{(0)}}^2\left(h_{AB}h^{AB}\right)+\mathcal{O}(h^3) \, .
\end{align}
We now impose the gauge condition $\nabla^Ah_{AB}=0$.
Then the action has the following form,
\begin{align}
S = \int d^5 x \sqrt{-g^{(0)}}\biggl[\frac{1}{2\kappa^2}
\biggl(&R^{(0)} +\frac{1}{4}h^{AB}\nabla^{(0)2}h_{AB}
 -\frac{1}{4}R^{(0)}h_{AB}h^{AB}\nn 
&+\frac{1}{2}R^{(0)}_{AB}h^{AD}h^B_D
+\frac{1}{2}R^{(0)ACBD}h_{AB}h_{CD}\biggr)
 -\frac{1}{4}h_{AB}h^{AB}\mathcal{L}_\mathrm{m}\biggr] \, ,
\end{align}
Then by the variation of the action with respect to $h_{AB}$, 
we obtain the following equation, 
\be
\label{Eq1}
0=\frac{1}{2\kappa^2}\left(
\frac{1}{2}\nabla^{(0)2}h^{AB}-\frac{1}{2}R^{(0)}h^{AB}
+\frac{1}{2}R^{(0)AC}h^B_C 
+\frac{1}{2}R^{(0)BC}h^A_C+R^{(0)ACBD}h_{CD}\right) 
 -\frac{1}{2}\mathcal{L}_\mathrm{m}h^{AB} \, .
\ee
Because we are interested in the $(5,\mu)$ component, we put 
$h_{\mu\nu}=h_{55}=0$. 
Then $(5,\mu)$ component of Eq.~(\ref{Eq1}) has the following form, 
\be
0=\frac{1}{2\kappa^2}\left(
\nabla^{(0)2}h^{5\mu}-R^{(0)}h^{5\mu }+R^{(0)55}h^\mu_5 
+R^{(0)\mu\nu}h^5_\nu-2R^{(0)\mu5\nu5}h_{5\nu}\right)
 -\mathcal{L}_\mathrm{m}h^{5\mu } \, .
\label{215830_25Nov16}
\ee

\subsection{Localization on Flat Domain Wall}

Before considering the FRW universe, we first consider the case that 
the four dimensional domain wall is flat as in the Randall-Sundrum model. 
Then the metric has the following form, 
\begin{align}
ds^2=\e^{u(w)}\eta_{\mu\nu}dx^\mu dx^\nu+dw^2\, .
\end{align}
Then Eq.~(\ref{215830_25Nov16}) has the following form,
\be
\left[\frac{1}{2\kappa^2}\left(\e^{-u}\partial_\nu\partial^\nu+\partial^2_5
+u^\prime\partial_5+2u^{\prime\prime}+u^{\prime2}\right)
 -\mathcal{L}_\mathrm{m}\right]h_{5\mu}=0 \, .
\label{221221_25Nov16}
\ee
The deviation of (\ref{221221_25Nov16}) is given in the Appendix 
\ref{AppendixA}. 
By assuming $h_{5\mu}(x^\nu,w)=N(w)A_\mu(x^\nu)$, 
we consider the following Lagrangian density of the scalar field 
instead of (\ref{pc1})  
(see Ref.~\cite{Toyozato:2012zh}), 
\be
\label{dw1}
\mathcal{L}_\mathrm{m}=-\frac{1}{2}\mathcal{C}(\chi)
\partial_A\chi \partial^A\chi-\mathcal{V}(\chi) 
=\frac{3}{2}u^{\prime\prime}+\frac{3}{2}u^{\prime2} \, .
\ee
The Lagrangian density is given by putting 
$A (\phi,\chi) = B (\phi,\chi) =0$, 
$\mathcal{C}(\chi) = \left. C (\phi,\chi) \right|_{\phi=0}$, 
and $\mathcal{V}(\chi) = \left. V (\phi,\chi) \right|_{\phi=0}$. 
We also used (\ref{pc7}) in the second equality in (\ref{dw1}). 
Then we obtain, 
\begin{align}
\left(N\e^{-u}\partial_\nu\partial^\nu+N^{\prime\prime}
+N^\prime u^\prime-Nu^{\prime\prime}-2Nu^{\prime2}\right)A_\mu=0 \, . 
\label{151324_19Nov16}
\end{align}
If we choose 
\begin{align}
\label{Neu}
N\propto \e^{u}\ , ,
\end{align}
Eq.~(\ref{151324_19Nov16}) coincides with the expression of the standard 
equation for the vector field in four dimensions, 
\begin{align}
\partial_\nu\partial^\nu A_\mu=0 \, .
\end{align}
For example, we consider the case, $u(w)=-2\sqrt{w^2+w^2_0}$, 
we find $N\propto \e^{-2\sqrt{w^2+w^2_0}}\to \e^{-2|w|}\ (w_0\to0)$ 
and therefore there occurs the localization of the vector field. 
If the number of the extra dimensions is not one but there are several 
extra dimensions and furthermore if the extra dimensions have a structure 
of the non-ablian group, there could appear the non-abelian gauge theory 
localized on the domain wall. 

\subsection{Localization on the Dynamical Domain Wall}

We now consider the case that the domain wall is dynamical, that is, 
the FRW universe is embedded in five dimensional bulk space-time 
as in (\ref{metricR}). 
Then the equation for the graviton is given by 
\begin{align}
0=&\frac{1}{2\kappa^2}\biggl\{\left(\partial^2_5+2u^{\prime\prime}
+u^\prime\partial_5+u^{\prime2}\right)h_{5\mu}\nn 
&+\e^{-u}\biggl[\left(\hat{\nabla}^2-2\ddot{u}-\frac{4\dot{a}\dot{u}}{a}
 -\frac{\dot{u}^2}{2}-\frac{6\ddot{a}}{a}-\frac{6\dot{a}^2}{a^2}\right)h_{5\mu}
+\left(\left(\ddot{u}+\frac{\dot{u}\dot{a}}{a}\right)h_{50}
 -\dot{u}\hat{g}^{\alpha\gamma}\partial_\alpha h_{5\gamma}\right)
\delta^0_\mu\nn 
&-\dot{u}\partial_\mu h_{50}\biggr]\biggr\}-\mathcal{L}_\mathrm{m}h_{5\mu}
\, .
\label{170308_23Nov16}
\end{align}
The derivation of (\ref{170308_23Nov16}) is given in Appendix \ref{AppendixB}. 
We assume the Lagrangian density $\mathcal{L}_\mathrm{m}$ is given by 
(\ref{pc1}). 
By substituting the expression of $\mathcal{L}_\mathrm{m}$  
into (\ref{170308_23Nov16}) and using the gauge fixing condition 
$\nabla^Ah_{A5}=0$, again, we find 
\be
\label{Eq64}
0= \left[(\partial^2_5+u^\prime\partial_5-u^{\prime\prime}-2u^{\prime2})
+\e^{-u}\left(\hat{\nabla}^2-\frac{2\ddot{a}}{a}-\frac{4\dot{a}^2}{a^2}\right)
\right]h_{5\mu} 
+\e^{-u}\left(\ddot{u}+\frac{\dot{u}^2}{a}
+\frac{4\dot{u}\dot{a}}{a}\right)h_{50}\delta^0_\mu
 -\e^{-u}\dot{u}\partial_\mu h_{50} \, .
\ee
By assuming $h_{5\mu}(x^\nu,w)=N(w)A_\mu(x^\nu)$, Eq.~(\ref{Eq64}) 
can be rewritten as 
\begin{align}
\label{Eq65}
0=&\left(N^{\prime\prime}+N^\prime u^\prime
 -Nu^{\prime\prime}-2Nu^{\prime2}\right)A_\mu 
+N\e^{-u}\biggl[\left(\hat{\nabla}^2-\frac{2\ddot{a}}{a}
 -\frac{4\dot{a}^2}{a^2}\right)A_\mu+\left(\ddot{u}+\frac{\dot{u}^2}{4}
+\frac{4\dot{u}\dot{a}}{a}\right)A_0\delta^0_\mu
 -\dot{u}\partial_\mu A_0\biggr] \nn 
=&\left(N^{\prime\prime}+N^\prime u^\prime
 -Nu^{\prime\prime}-2Nu^{\prime2}\right)A_\mu 
+N\e^{-u}\biggl[\left(\hat{\nabla}^2-\frac{2\ddot{a}}{a}
 -\frac{4\dot{a}^2}{a^2}-\frac{\dot{a}\dot{u}}{a}\right)A_\mu+\left(\ddot{u}
+\frac{\dot{u}^2}{4}+\frac{5\dot{u}\dot{a}}{a}\right)A_0\delta^0_\mu
 -\dot{u}\hat{\nabla}_\mu A_0\biggr]  \, .
\end{align}
In case that $u$ can be separated into a sum of $w$-dependent part 
$u_w(w)$ 
and $t$-dependent part $u_t(t)$, that is, $u(w,t)=u_w(w) + u_t (t)$ as 
in (\ref{metricRB})  
$\left( W(w)\propto \e^{u_w(w)}\, , \ T(t) \propto \e^{u_t(t)}\right)$, 
if we choose $N(w) \propto \e^{u_w(w)}$ 
as in (\ref{Neu}), the first term vanishes in (\ref{Eq65}) and we obtain
\be
\label{Eq65B}
0=\left(\hat{\nabla}^2-\frac{2\ddot{a}}{a}-\frac{4\dot{a}^2}{a^2}
 -\frac{\dot{a}\dot{u_t}}{a}\right)A_\mu+\left(\ddot{u_t}
+\frac{\dot{u_t}^2}{4}+\frac{5\dot{u_t}\dot{a}}{a}\right)A_0\delta^0_\mu
 -\dot{u_t}\hat{\nabla}_\mu A_0 \, .
\ee
On the other hand, the vector field in the four dimensional FRW space-time 
obeys the following equation, 
\begin{align}
\label{Eq67}
0=&\hat{\nabla}^2A_\mu-\hat{\nabla}_\nu\hat{\nabla}_\mu A^\nu
=\hat{\nabla}^2A_\mu-\hat{\nabla}_\mu\hat{\nabla}_\nu A^\nu
 -\hat{R}_{\lambda\mu}A^\lambda\nn 
=&\hat{\nabla}^2A_\mu-\hat{\nabla}_\mu\hat{\nabla}_\nu A^\nu
 -\left(\frac{\ddot{a}}{a}+\frac{2\dot{a}^2}{a^2}\right)A_\mu
 -\left(\frac{2\ddot{a}}{a}-\frac{2\dot{a}^2}{a^2}\right)A_0\delta^0_\mu \, .
\end{align}
There are some differences between (\ref{Eq65B}) and (\ref{Eq67}) 
even if we choose the gauge condition $\hat{\nabla}_\nu A^\nu=0$. 
Therefore the vector field can localized on the domain wall even if 
dynamical but there appear some corrections from the extra dimensions. 

\section{Summary \label{VI}}

In summary, 
the localization of vector field in the model 
\cite{Toyozato:2012zh, Higuchi:2014bya} has been investigated by using 
three formulations. 
\begin{enumerate}
\item The first formulation was just the review of the work in 
\cite{Toyozato:2015xfa}, where we have used the action of the five 
dimensional vector field. 
\item The second formulation was an extension of those 
in \cite{Dvali:1996xe} 
and \cite{Dvali:1996xe}. In this formulation, the non-abelian gauge field is 
confined in the bulk space-time but massless $U(1)$ gauge field appears 
due to the spontaneous breakdown of the gauge symmetry. 
In \cite{Davies:2007xr}, the confinement in the bulk space-time was assumed 
in the five dimensional bulk space-time. 
It is known, however that there could be a phase transition in the 
dimensions higher than four and the confinement could occur only in 
the strong coupling region. 
Then we may consider the model where the scalar field plays a role of the 
gauge coupling. 
The strong coupling phase can be always realized if the gauge coupling is 
given by the scalar field depending on the corrdinate in the extra dimension 
and the coupling becomes strong enough 
and the confinement always occurs 
in the bulk space-time. 
\item The third formulation was given by the Kaluza-Klein modes coming 
from the five dimensional graviton. 
In the second Randall-Sundrum model \cite{Randall:1999vf}, the graviton was 
localized on the brane. 
The localized graviton can be regarded as a zero mode 
of the five dimensional graviton. 
We have shown that the $(5,\mu)$ components 
$\left(\mu=0,1,2,3\right)$ of the graviton are also localized on the domain 
wall and can be regarded as the vector field on the four dimensional domain 
wall. 
We found that, however, some corrections appear from the bulk extra 
dimension if we consider the dynamical domain wall. 
An interesting point is that if we have several extra dimensions and the 
extra dimensions have a symmetry under the non-ablian group 
transformation, there could appear the non-abelian gauge theory 
localized on the domain wall. 
\end{enumerate}
Then it might be interesting to realize the GUT on the domain wall. 

\section*{Acknowledgments.}

This work is supported (in part) by 
MEXT KAKENHI Grant-in-Aid for Scientific Research on Innovative Areas ``Cosmic
Acceleration''  (No. 15H05890) and the JSPS Grant-in-Aid for Scientific Research (C) \# 23540296 (S.N.).

\appendix

\section{Derivation of (\ref{221221_25Nov16}) \label{AppendixA}}

Here we consider the derivation of of (\ref{221221_25Nov16}). 
We have the following expressions of the connection 
\begin{align}
\Gamma^5_{\mu\nu}=-\frac{u^\prime}{2}\e^u\eta_{\mu\nu}\, ,\quad 
\Gamma^\mu_{\nu5}=\frac{u^\prime}{2}\delta^\mu_\nu \, ,
\end{align}
the Riemann tensor, 
\begin{align}
R^{(0)}_{\mu5\nu5}=-\e^u\left(\frac{1}{2}u^{\prime\prime}
+\frac{1}{4}u^{\prime2}\right)\eta_{\mu\nu} \, ,
\end{align}
Ricci tensor, 
\begin{align}
 R^{(0)}_{\mu\nu}=-\e^u\left(\frac{u^{\prime\prime}}{2}
 +u^{\prime2}\right)\eta_{\mu\nu},\ R^{(0)}_{55}
 =-2u^{\prime\prime}-u^{\prime2} \, ,
\end{align}
and the scalar curvature 
\begin{align}
 R^{(0)}=-4u^{\prime\prime}-5u^{\prime2} \, .
\end{align}
Then we find 
\begin{align}
g^{AB}\nabla^{(0)}_A & \nabla^{(0)}_Bh_{5\mu} \nn
=&g^{AB}\left(\partial_A\nabla^{(0)}_Bh_{5\mu}-\Gamma^C_{AB}
\nabla^{(0)}_Ch_{5\mu}-\Gamma^C_{A5}\nabla^{(0)}_Bh_{C\mu}
 -\Gamma^C_{A\mu}\nabla^{(0)}_Bh_{5C}\right)\nn 
 =&g^{AB}\partial_A\left(\partial_Bh_{5\mu}-\Gamma^C_{B5}h_{C\mu}
 -\Gamma^C_{B\mu}h_{5C}\right) 
 -g^{AB}\Gamma^C_{AB}\left(\partial_Ch_{5\mu}-\Gamma^D_{C5}h_{D\mu}
 -\Gamma^D_{C\mu}h_{5D}\right) \nn 
&-g^{AB}\Gamma^C_{A5}\left(\partial_Bh_{C\mu}-\Gamma^D_{BC}h_{D\mu}
 -\Gamma^D_{B\mu}h_{CD}\right) \nn 
&-g^{AB}\Gamma^C_{A\mu}\left(\partial_Bh_{5C}-\Gamma^D_{B5}h_{DC}
 -\Gamma^D_{BC}h_{5D}\right) \nn
=&g^{AB}\partial_A\partial_Bh_{5\mu}-g^{AB}\partial_A\Gamma^C_{B5}
h_{C\mu}-2g^{AB}\Gamma^C_{B5}\partial_Ah_{C\mu}
 -g^{AB}\partial_A\Gamma^C_{B\mu}h_{5C}
 -2g^{AB}\Gamma^C_{B\mu}\partial_Ah_{5C} \nn 
&-g^{AB}\Gamma^C_{AB}\partial_Ch_{5\mu}
+g^{AB}\Gamma^C_{AB}\Gamma^D_{C5}h_{D\mu}
+g^{AB}\Gamma^C_{AB}\Gamma^D_{C\mu}h_{5D} \nn 
&+g^{AB}\Gamma^C_{A5}\Gamma^D_{BC}h_{D\mu}
+2g^{AB}\Gamma^C_{A5}\Gamma^D_{B\mu}h_{CD}
+g^{AB}\Gamma^C_{A\mu}\Gamma^D_{BC}h_{5D} \nn 
 =&\partial^2_5h_{5\mu}
+\e^{-u}\hat{g}^{\alpha\beta}\partial_\alpha\partial_\beta h_{5\mu}
 -\partial_5\Gamma^5_{55}h_{5\mu}
 -\e^{-u}\hat{g}^{\alpha\beta}\partial_\alpha\Gamma^5_{\beta5}
h_{5\mu}-2\Gamma^5_{55}\partial_5h_{5\mu}
 -2\e^{-u}\hat{g}^{\alpha\beta}\Gamma^5_{\beta5}
\partial_\alpha h_{5\mu} \nn 
& -\partial_5\Gamma^\gamma_{5\mu}h_{5\gamma}
 -\e^{-u}\hat{g}^{\alpha\beta}\partial_\alpha\Gamma^\gamma_{\beta\mu}
h_{5\gamma}-2\Gamma^\gamma_{5\mu}\partial_5h_{5\gamma}
 -2\e^{-u}\hat{g}^{\alpha\beta}\Gamma^\gamma_{\beta\mu}
\partial_\alpha h_{5\gamma} \nn 
&-\Gamma^C_{55}\partial_Ch_{5\mu}
 -\e^{-u}\hat{g}^{\alpha\beta}\Gamma^C_{\alpha\beta}\partial_Ch_{5\mu}
+\Gamma^C_{55}\Gamma^5_{C5}h_{5\mu}
+\e^{-u}\hat{g}^{\alpha\beta}\Gamma^C_{\alpha\beta}\Gamma^5_{C5}h_{5\mu}
+\Gamma^C_{55}\Gamma^\delta_{C\mu}h_{5\delta}\nn 
& +\e^{-u}\hat{g}^{\alpha\beta}\Gamma^C_{\alpha\beta}
\Gamma^\delta_{C\mu}h_{5\delta} \nn 
&+\Gamma^C_{55}\Gamma^5_{5C}h_{5\mu}
+\e^{-u}\hat{g}^{\alpha\beta}\Gamma^C_{\alpha5}\Gamma^5_{\beta C}
h_{5\mu}+2\Gamma^C_{55}\Gamma^D_{5\mu}h_{CD}
+2\e^{-u}\hat{g}^{\alpha\beta}\Gamma^C_{\alpha5}\Gamma^D_{\beta\mu}
h_{CD}+\Gamma^C_{5\mu}\Gamma^\delta_{5C}h_{5\delta}\nn 
& +\e^{-u}\hat{g}^{\alpha\beta}\Gamma^C_{\alpha\mu}
\Gamma^\delta_{\beta C}h_{5\delta} \nn 
=&\partial^2_5h_{5\mu}-\partial_5\Gamma^\gamma_{5\mu}
h_{5\gamma}-2\Gamma^\gamma_{5\mu}\partial_5h_{5\gamma}
+\Gamma^C_{5\mu}\Gamma^\delta_{5C}h_{5\delta} \nn 
&+\e^{-u}\hat{g}^{\alpha\beta}\biggl[\partial_\alpha\partial_\beta h_{5\mu}
 -\partial_\alpha\Gamma^\gamma_{\beta\mu}h_{5\gamma}
 -2\Gamma^\gamma_{\beta\mu}\partial_\alpha h_{5\gamma}
 -\Gamma^C_{\alpha\beta}\partial_Ch_{5\mu}
+\Gamma^C_{\alpha\beta}\Gamma^\delta_{C\mu}h_{5\delta}
+\Gamma^C_{\alpha5}\Gamma^5_{\beta C}h_{5\mu}\nn  
&+2\Gamma^C_{\alpha5}\Gamma^D_{\beta\mu}h_{CD}
+\Gamma^C_{\alpha\mu}\Gamma^\delta_{\beta C}h_{5\delta}\biggr] \nn 
=&\partial^2_5h_{5\mu}-\frac{u^{\prime\prime}}{2}h_{5\mu}
 -u^\prime\partial_5h_{5\mu}+\frac{u^{\prime2}}{4}h_{5\mu} \nn 
&+\e^{-u}\eta^{\alpha\beta}\biggl[\partial_\alpha\partial_\beta h_{5\mu}
+\frac{u^\prime}{2}\e^u\eta_{\alpha\beta}\partial_5h_{5\mu}
 -\frac{u^{\prime2}}{4}\e^u\eta_{\alpha\beta}h_{5\mu}
 -\frac{u^{\prime2}}{4}\e^u\eta_{\alpha\beta}h_{5\mu}\nn 
&-\frac{u^{\prime2}}{2}\e^u\eta_{\beta\mu}h_{5\alpha}
 -\frac{u^{\prime2}}{4}\e^u\eta_{\alpha\mu}h_{5\beta}\biggr] \nn 
=&\left(\partial^2_5-\frac{u^{\prime\prime}}{2}+u^\prime\partial_5
 -\frac{5u^{\prime2}}{2}\right)h_{5\mu}
+\e^{-u}\eta^{\alpha\beta}\partial_\alpha\partial_\beta h_{5\mu} \, , \\
 -R^{(0)}h_{5\mu} & 
+R^{(0)55}h_{5\mu}+R^{(0)\ \nu}_{\ \ \ \mu}h_{5\nu}
 -2R^{(0)\ 5\nu5}_{\ \ \ \mu}h_{5\nu}\nn 
=&-(-4u^{\prime\prime}-5u^{\prime2})h_{5\mu}+(-2u^{\prime\prime}
 -u^{\prime2})h_{5\mu}-\left(\frac{u^{\prime\prime}}{2}
+u^{\prime2}\right)\delta^\nu_\mu h_{5\nu}
 -2\left(-\frac{u^{\prime\prime}}{2}
 -\frac{u^{\prime2}}{4}\right)\delta^\nu_\mu h_{5\nu} \nn 
=&\left(\frac{5u^{\prime\prime}}{2}+\frac{7u^{\prime2}}{2}\right)h_{5\mu} 
\end{align}
Then by substituting the above expressions into (\ref{215830_25Nov16}), 
we obtain Eq.~(\ref{221221_25Nov16}). 

\section{Derivation of (\ref{170308_23Nov16}) \label{AppendixB}}

We now consider the derivation of (\ref{170308_23Nov16}). 
The expressions of the connection and curvatures are given by
\begin{align}
& \Gamma^\mu_{\nu0}=\left(\frac{\dot{a}}{a}
+\frac{\dot{u}}{2}\right)\delta^\mu_\nu
 -\frac{\dot{a}}{a}\delta^\mu_0\delta^0_\nu \, , \quad 
\Gamma^0_{ij}=\left(\frac{\dot{a}}{a}+\frac{\dot{u}}{2}\right)\hat{g}_{ij} \, , 
\quad \Gamma^\mu_{\nu5}=\frac{u^\prime}{2}\delta^\mu_\nu \, , \quad 
\Gamma^5_{\mu\nu}=-\e^u\frac{u^\prime}{2}\hat{g}_{\mu\nu} \, , \\
&R^{(0)}=-4u^{\prime\prime}-5u^{\prime2}+3\e^{-u}\left(\ddot{u}
+\frac{\dot{u}^2}{2}+\frac{3\dot{a}\dot{u}}{a}+\frac{2\ddot{a}}{a}
+\frac{2\dot{a}^2}{a^2}\right) \, , \quad 
R^{(0)55}=-2u^{\prime\prime}-u^{\prime2} \, ,\nn
&R^{(0)}_{00}=\left[-\e^u\left(\frac{u^{\prime\prime}}{2}+u^{\prime2}\right)
+\frac{3\ddot{u}}{2}+\frac{3\dot{a}\dot{u}}{2a}\right]\hat{g}_{00}
+\hat{R}_{00} \, , \quad 
R^{(0)}_{ij}=\left[-\e^{u}\left(\frac{u^{\prime\prime}}{2}+u^{\prime2}\right)
+\frac{\ddot{u}}{2}+\frac{\dot{u}^2}{2}
+\frac{5\dot{a}\dot{u}}{2a}\right]\hat{g}_{ij}+\hat{R}_{ij}\, , \nn
&R^{(0)}_{\mu\nu}=\left[-\e^u\left(\frac{u^{\prime\prime}}{2}
+u^{\prime2}\right)+\frac{\ddot{u}}{2}+\frac{\dot{u}^2}{2}
+\frac{5\dot{a}\dot{u}}{2a}\right]\hat{g}_{\mu\nu}
+\left(\ddot{u}-\frac{\dot{u}^2}{2}
 -\frac{\dot{a}\dot{u}}{a}\right)\hat{g}_{00}\delta^0_\mu\delta^0_\nu
+\hat{R}_{\mu\nu} \, , \nn
&R^{(0)\mu5\nu5}=-\left(\frac{u^{\prime\prime}}{2}
+\frac{u^{\prime2}}{4}\right)g^{\mu\nu} \, .
\end{align}
We also find 
\begin{align}
\nabla^{(0)2}h_{5\mu}=&g^{AB}\left(\partial_A\nabla^{(0)}_Bh_{5\mu}
 -\Gamma^C_{AB}\nabla^{(0)}_Ch_{5\mu}
 -\Gamma^C_{A5}\nabla^{(0)}_Bh_{C\mu}
 -\Gamma^C_{A\mu}\nabla^{(0)}_Bh_{5C}\right) \nn
=&g^{AB}\partial_A\left(\partial_Bh_{5\mu}-\Gamma^C_{B5}h_{C\mu}
 -\Gamma^C_{B\mu}h_{5C}\right)  
 -g^{AB}\Gamma^C_{AB}\left(\partial_Ch_{5\mu}-\Gamma^D_{C5}h_{D\mu}
 -\Gamma^D_{C\mu}h_{5D}\right)  \nn
 &-g^{AB}\Gamma^C_{A5}\left(\partial_Bh_{C\mu}-\Gamma^D_{BC}h_{D\mu}
 -\Gamma^D_{B\mu}h_{CD}\right)  
 -g^{AB}\Gamma^C_{A\mu}\left(\partial_Bh_{5C}-\Gamma^D_{B5}h_{DC}
 -\Gamma^D_{BC}h_{5D}\right) \nn
=&g^{AB}\partial_A\partial_Bh_{5\mu}
 -g^{AB}\partial_A\Gamma^C_{B5}h_{C\mu}
 -2g^{AB}\Gamma^C_{B5}\partial_Ah_{C\mu}
 -g^{AB}\partial_A\Gamma^C_{B\mu}h_{5C}
 -2g^{AB}\Gamma^C_{B\mu}\partial_Ah_{5C}  \nn
&-g^{AB}\Gamma^C_{AB}\partial_Ch_{5\mu}
+g^{AB}\Gamma^C_{AB}\Gamma^D_{C5}h_{D\mu}
+g^{AB}\Gamma^C_{AB}\Gamma^D_{C\mu}h_{5D}  \nn
&+g^{AB}\Gamma^C_{A5}\Gamma^D_{BC}h_{D\mu}
+2g^{AB}\Gamma^C_{A5}\Gamma^D_{B\mu}h_{CD}
+g^{AB}\Gamma^C_{A\mu}\Gamma^D_{BC}h_{5D}  \nn
=&\partial^2_5h_{5\mu}
+\e^{-u}\hat{g}^{\alpha\beta}\partial_\alpha\partial_\beta h_{5\mu}
 -\partial_5\Gamma^5_{55}h_{5\mu}
 -\e^{-u}\hat{g}^{\alpha\beta}\partial_\alpha\Gamma^5_{\beta5}
h_{5\mu}
 -2\Gamma^5_{55}\partial_5h_{5\mu}
 -2\e^{-u}\hat{g}^{\alpha\beta}\Gamma^5_{\beta5}\partial_\alpha h_{5\mu}  \nn
& -\partial_5\Gamma^\gamma_{5\mu}h_{5\gamma}
-\e^{-u}\hat{g}^{\alpha\beta}\partial_\alpha
\Gamma^\gamma_{\beta\mu}h_{5\gamma}
 -2\Gamma^\gamma_{5\mu}\partial_5h_{5\gamma}
 -2\e^{-u}\hat{g}^{\alpha\beta}\Gamma^\gamma_{\beta\mu}
\partial_\alpha h_{5\gamma}  \nn
&-\Gamma^C_{55}\partial_Ch_{5\mu}
 -\e^{-u}\hat{g}^{\alpha\beta}\Gamma^C_{\alpha\beta}\partial_Ch_{5\mu}
 +\Gamma^C_{55}\Gamma^5_{C5}h_{5\mu}
 +\e^{-u}\hat{g}^{\alpha\beta}\Gamma^C_{\alpha\beta}\Gamma^5_{C5}
 h_{5\mu}+\Gamma^C_{55}\Gamma^\delta_{C\mu}h_{5\delta} \nn
&+\e^{-u}\hat{g}^{\alpha\beta}\Gamma^C_{\alpha\beta}
\Gamma^\delta_{C\mu}h_{5\delta}  \nn
&+\Gamma^C_{55}\Gamma^5_{5C}h_{5\mu}
+\e^{-u}\hat{g}^{\alpha\beta}\Gamma^C_{\alpha5}\Gamma^5_{\beta C}
h_{5\mu}
+2\Gamma^C_{55}\Gamma^D_{5\mu}h_{CD}
+2\e^{-u}\hat{g}^{\alpha\beta}\Gamma^C_{\alpha5}\Gamma^D_{\beta\mu}
h_{CD}
+\Gamma^C_{5\mu}\Gamma^\delta_{5C}h_{5\delta} \nn
&+\e^{-u}\hat{g}^{\alpha\beta}\Gamma^C_{\alpha\mu}
\Gamma^\delta_{\beta C}h_{5\delta}  \nn
=&\partial^2_5h_{5\mu}-\partial_5\Gamma^\gamma_{5\mu}
h_{5\gamma}-2\Gamma^\gamma_{5\mu}\partial_5h_{5\gamma}
+\Gamma^C_{5\mu}\Gamma^\delta_{5C}h_{5\delta}  \nn
&+\e^{-u}\hat{g}^{\alpha\beta}\biggl[\partial_\alpha\partial_\beta h_{5\mu}
 -\partial_\alpha\Gamma^\gamma_{\beta\mu}h_{5\gamma}
 -2\Gamma^\gamma_{\beta\mu}\partial_\alpha h_{5\gamma}
 -\Gamma^C_{\alpha\beta}\partial_Ch_{5\mu}
+\Gamma^C_{\alpha\beta}\Gamma^\delta_{C\mu}h_{5\delta}
+\Gamma^C_{\alpha5}\Gamma^5_{\beta C}h_{5\mu} \nn 
&+2\Gamma^C_{\alpha5}\Gamma^D_{\beta\mu}h_{CD}
+\Gamma^C_{\alpha\mu}\Gamma^\delta_{\beta C}h_{5\delta}\biggr] \nn
=&\partial^2_5h_{5\mu}-\partial_5\Gamma^\gamma_{5\mu}
h_{5\gamma}-2\Gamma^\gamma_{5\mu}\partial_5h_{5\gamma}
+\Gamma^C_{5\mu}\Gamma^\delta_{5C}h_{5\delta}  \nn
&+\e^{-u}\biggl[\hat{g}^{\alpha\beta}\partial_\alpha\partial_\beta h_{5\mu}
 -(\mathrm{\ref{152218_23Nov16}})-2(\mathrm{\ref{152305_23Nov16}})
 -(\mathrm{\ref{152414_23Nov16}}) +(\mathrm{\ref{152509_23Nov16}})
+(\mathrm{\ref{152747_23Nov16}}) +2(\mathrm{\ref{152649_23Nov16}})
+(\mathrm{\ref{152838_23Nov16}})\biggr] \, .
\end{align}
We now find the explicit forms of Eqs.~(\ref{152218_23Nov16}), 
(\ref{152305_23Nov16}), (\ref{152414_23Nov16}), (\ref{152509_23Nov16}), 
(\ref{152747_23Nov16}), (\ref{152649_23Nov16}), and (\ref{152838_23Nov16}). 
Because 
\begin{align}
\Gamma^\alpha_{\beta\gamma}
=&\frac{1}{2}g^{\alpha M}\left(g_{M\beta,\gamma}
+g_{M\gamma,\beta}-g_{\beta\gamma,M}\right) \nn
=&\frac{1}{2}\hat{g}^{\alpha\mu}\left(\hat{g}_{\mu\beta,\gamma}
+\hat{g}_{\mu\gamma,\beta}-\hat{g}_{\beta\gamma,\mu}
+\hat{g}_{\mu\beta}\partial_\gamma u
+\hat{g}_{\mu\gamma}\partial_\beta u
 -\hat{g}_{\beta\gamma}\partial_\mu u\right)  \nn
=&\hat{\Gamma}^\alpha_{\beta\gamma}
+\frac{1}{2}\left(\delta^\alpha_\beta\partial_\gamma u
+\delta^\alpha_\gamma\partial_\beta u
 -\hat{g}_{\beta\gamma}\hat{g}^{\alpha\delta}\partial_\delta u\right) \, , 
\end{align}
we obtain 
\begin{align}
\hat{g}^{\alpha\beta}\partial_\alpha\Gamma^\gamma_{\beta\mu}h_{5\gamma}
=&\hat{g}^{\alpha\beta}\partial_\alpha\left[\hat{\Gamma}^\gamma_{\beta\mu}
+\frac{1}{2}\left(\delta^\gamma_\beta\partial_\mu u
+\delta^\gamma_\mu\partial_\beta u
 -\hat{g}_{\beta\mu}\hat{g}^{\gamma\delta}\partial_\delta u\right)
\right]h_{5\gamma} \nn
=&\hat{g}^{\alpha\beta}\partial_\alpha\hat{\Gamma}^\gamma_{\beta\mu}
h_{5\gamma}
+\frac{1}{2}(\partial_\alpha\partial_\mu u)\hat{g}^{\alpha\gamma}h_{5\gamma}
+\frac{1}{2}\hat{g}^{\alpha\beta}(\partial_\alpha\partial_\beta u)h_{5\mu} \nn
&-\frac{1}{2}\hat{g}^{\alpha\beta}(\partial_\alpha\hat{g}_{\beta\mu})
(\partial_\delta u)\hat{g}^{\gamma\delta}h_{5\gamma}
 -\frac{1}{2}(\partial_\mu\hat{g}^{\gamma\delta})
(\partial_\delta u)h_{5\gamma}
 -\frac{1}{2}(\partial_\mu\partial_\delta u)\hat{g}^{\gamma\delta}
h_{5\gamma}  \nn
=&\hat{g}^{\alpha\beta}\partial_\alpha\hat{\Gamma}^\gamma_{\beta\mu}
h_{5\gamma}
+\frac{1}{2}\hat{g}^{\alpha\beta}(\partial_\alpha\partial_\beta u)h_{5\mu}
 -\frac{1}{2}\hat{g}^{\alpha\beta}(\partial_\alpha\hat{g}_{\beta\mu})
(\partial_\delta u)\hat{g}^{\gamma\delta}h_{5\gamma}
 -\frac{1}{2}(\partial_\mu\hat{g}^{\gamma\delta})
(\partial_\delta u)h_{5\gamma}   \nn
=&\hat{g}^{\alpha\beta}\partial_\alpha\hat{\Gamma}^\gamma_{\beta\mu}
h_{5\gamma}
 -\frac{1}{2}\ddot{u}h_{5\mu} \, , 
\label{152218_23Nov16} \\
\hat{g}^{\alpha\beta}\Gamma^\gamma_{\beta\mu}\partial_\alpha h_{5\gamma}
=&\hat{g}^{\alpha\beta}\left[\hat{\Gamma}^\gamma_{\beta\mu}
+\frac{1}{2}\left(\delta^\gamma_\beta\partial_\mu u
+\delta^\gamma_\mu\partial_\beta u
 -\hat{g}_{\beta\mu}\hat{g}^{\gamma\delta}\partial_\delta u\right)\right]
\partial_\alpha h_{5\gamma} \nn
=&\hat{g}^{\alpha\beta}\hat{\Gamma}^\gamma_{\beta\mu}\partial_\alpha 
h_{5\gamma}
+\frac{1}{2}(\partial_\mu u)\hat{g}^{\alpha\gamma}\partial_\alpha 
h_{5\gamma}
+\frac{1}{2}\hat{g}^{\alpha\beta}(\partial_\beta u)\partial_\alpha h_{5\mu}
 -\frac{1}{2}(\partial_\delta u)\hat{g}^{\gamma\delta}\partial_\mu 
h_{5\gamma}  \nn
=&\hat{g}^{\alpha\beta}\hat{\Gamma}^\gamma_{\beta\mu}\partial_\alpha 
h_{5\gamma}
+\frac{1}{2}(\partial_\mu u)\hat{g}^{\alpha\gamma}\partial_\alpha 
h_{5\gamma}
 -\frac{1}{2}\dot{u}\partial_0h_{5\mu}
+\frac{1}{2}\dot{u}\partial_\mu h_{50} \, , 
\label{152305_23Nov16} \\
\hat{g}^{\alpha\beta}\Gamma^C_{\alpha\beta}\partial_Ch_{5\mu}
=&\hat{g}^{\alpha\beta}\left[\hat{\Gamma}^\gamma_{\alpha\beta}
+\frac{1}{2}\left(\delta^\gamma_\alpha\partial_\beta u
+\delta^\gamma_\beta\partial_\alpha u
 -\hat{g}_{\alpha\beta}\hat{g}^{\gamma\delta}\partial_\delta u\right)\right]
\partial_\gamma h_{5\mu}+\hat{g}^{\alpha\beta}\Gamma^5_{\alpha\beta}
\partial_5h_{5\mu} \nn
=&\hat{g}^{\alpha\beta}\hat{\Gamma}^\gamma_{\alpha\beta}\partial_\gamma 
h_{5\mu}
+\frac{1}{2}(\partial_\beta u)\hat{g}^{\beta\gamma}\partial_\gamma h_{5\mu}
+\frac{1}{2}(\partial_\alpha u)\hat{g}^{\gamma\alpha}\partial_\gamma 
h_{5\mu}
 -2(\partial_\delta u)\hat{g}^{\gamma\delta}\partial_\gamma h_{5\mu}
+\hat{g}^{\alpha\beta}\Gamma^5_{\alpha\beta}\partial_5h_{5\mu}  \nn
=&\hat{g}^{\alpha\beta}\hat{\Gamma}^\gamma_{\alpha\beta}
\partial_\gamma h_{5\mu}-(\partial_\alpha u)\hat{g}^{\alpha\beta}
\partial_\beta h_{5\mu}+\hat{g}^{\alpha\beta}\Gamma^5_{\alpha\beta}
\partial_5h_{5\mu}  \nn
=&\hat{g}^{\alpha\beta}\hat{\Gamma}^\gamma_{\alpha\beta}\partial_\gamma 
h_{5\mu}
+\dot{u}\partial_0h_{5\mu}-2\e^uu^\prime\partial_5h_{5\mu} \, , 
\label{152414_23Nov16} \\
\hat{g}^{\alpha\beta}\Gamma^C_{\alpha\beta}\Gamma^\delta_{C\mu}
h_{5\delta}=&\hat{g}^{\alpha\beta}\left[\hat{\Gamma}^\gamma_{\alpha\beta}
+\frac{1}{2}\left(\delta^\gamma_\alpha\partial_\beta u
+\delta^\gamma_\beta\partial_\alpha u
 -\hat{g}_{\alpha\beta}\hat{g}^{\gamma\lambda}\partial_\lambda u\right)
\right]\left[\hat{\Gamma}^\delta_{\gamma\mu}
+\frac{1}{2}\left(\delta^\delta_\gamma\partial_\mu u
+\delta^\delta_\mu\partial_\gamma u
 -\hat{g}_{\gamma\mu}\hat{g}^{\delta\rho}\partial_\rho u\right)
\right]h_{5\delta} \nn
&+\hat{g}^{\alpha\beta}\Gamma^5_{\alpha\beta}\Gamma^\delta_{5\mu}
h_{5\delta} \nn 
=&\hat{g}^{\alpha\beta}\hat{\Gamma}^\gamma_{\alpha\beta}
\hat{\Gamma}^\delta_{\gamma\mu}h_{5\delta}
+\frac{1}{2}\hat{g}^{\alpha\beta}\hat{\Gamma}^\gamma_{\alpha\beta}
(\partial_\mu u)h_{5\gamma}+\frac{1}{2}\hat{g}^{\alpha\beta}
\hat{\Gamma}^\gamma_{\alpha\beta}(\partial_\gamma u)h_{5\mu}
 -\frac{1}{2}\hat{g}^{\alpha\beta}\hat{\Gamma}^\gamma_{\alpha\beta}
\hat{g}_{\gamma\mu}(\partial_\rho u)\hat{g}^{\delta\rho}h_{5\delta}  \nn
&+\frac{1}{2}(\partial_\beta u)\hat{g}^{\beta\gamma}
\hat{\Gamma}^\delta_{\gamma\mu}h_{5\delta}
+\frac{1}{4}(\partial_\beta u)(\partial_\mu u)\hat{g}^{\beta\delta}h_{5\delta}
+\frac{1}{4}(\partial_\beta u)\hat{g}^{\beta\gamma}
(\partial_\gamma u)h_{5\mu}
 -\frac{1}{4}(\partial_\mu u)(\partial_\rho u)\hat{g}^{\delta\rho}h_{5\delta} \nn
&+\frac{1}{2}(\partial_\alpha u)\hat{g}^{\alpha\gamma}
\hat{\Gamma}^\delta_{\gamma\mu}h_{5\delta}
+\frac{1}{4}(\partial_\alpha u)(\partial_\mu u)\hat{g}^{\alpha\delta}h_{5\delta}
+\frac{1}{4}(\partial_\alpha u)\hat{g}^{\alpha\gamma}
(\partial_\gamma u)h_{5\mu}-\frac{1}{4}(\partial_\mu u)(\partial_\rho u)
\hat{g}^{\delta\rho}h_{5\delta}  \nn
&-2(\partial_\lambda u)\hat{g}^{\gamma\lambda}
\hat{\Gamma}^\delta_{\gamma\mu}h_{5\delta}
 -(\partial_\lambda u)(\partial_\mu u)\hat{g}^{\lambda\delta}h_{5\delta}
 -(\partial_\lambda u)\hat{g}^{\gamma\lambda}(\partial_\gamma u)h_{5\mu}
+(\partial_\mu u)(\partial_\rho u)\hat{g}^{\delta\rho}h_{5\delta}  \nn
&+\hat{g}^{\alpha\beta}\Gamma^5_{\alpha\beta}\Gamma^\delta_{5\mu}
h_{5\delta}  \nn
 =&\hat{g}^{\alpha\beta}\hat{\Gamma}^\gamma_{\alpha\beta}
\hat{\Gamma}^\delta_{\gamma\mu}h_{5\delta}
+\frac{1}{2}\hat{g}^{\alpha\beta}\hat{\Gamma}^\gamma_{\alpha\beta}
(\partial_\mu u)h_{5\gamma}+\frac{1}{2}\hat{g}^{\alpha\beta}
\hat{\Gamma}^\gamma_{\alpha\beta}(\partial_\gamma u)h_{5\mu}
 -\frac{1}{2}\hat{g}^{\alpha\beta}\hat{\Gamma}^\gamma_{\alpha\beta}
\hat{g}_{\gamma\mu}(\partial_\rho u)\hat{g}^{\delta\rho}h_{5\delta}  \nn
&-(\partial_\alpha u)\hat{g}^{\alpha\gamma}
\hat{\Gamma}^\delta_{\gamma\mu}h_{5\delta}
 -\frac{1}{2}(\partial_\alpha u)\hat{g}^{\alpha\beta}(\partial_\beta u)h_{5\mu}
+\hat{g}^{\alpha\beta}\Gamma^5_{\alpha\beta}\Gamma^\delta_{5\mu}
h_{5\delta}  \nn
=&\hat{g}^{\alpha\beta}\hat{\Gamma}^\gamma_{\alpha\beta}
\hat{\Gamma}^\delta_{\gamma\mu}h_{5\delta}
+\frac{3\dot{a}\dot{u}}{2a}h_{50}\delta^0_\mu
+\frac{3\dot{a}\dot{u}}{2a}h_{5\mu}
 -\frac{3\dot{a}\dot{u}}{2a}h_{50}\delta^0_\mu \nn
&+\frac{\dot{a}\dot{u}}{a}h_{5\mu}
 -\frac{\dot{a}\dot{u}}{a}h_{50}\delta^0_\mu+\frac{\dot{u}^2}{2}h_{5\mu}
 -\e^uu^{\prime2}h_{5\mu} \nn
=&\hat{g}^{\alpha\beta}\hat{\Gamma}^\gamma_{\alpha\beta}
\hat{\Gamma}^\delta_{\gamma\mu}h_{5\delta}
 -\frac{\dot{a}\dot{u}}{a}h_{50}\delta^0_\mu+\frac{5\dot{a}\dot{u}}{2a}h_{5\mu}
+\frac{\dot{u}^2}{2}h_{5\mu}-\e^uu^{\prime2}h_{5\mu} \, , 
\label{152509_23Nov16} \\
\hat{g}^{\alpha\beta}\Gamma^C_{\alpha5}\Gamma^5_{\beta C}h_{5\mu}
=&\hat{g}^{\alpha\beta}\Gamma^\gamma_{\alpha5}\Gamma^5_{\beta\gamma}
h_{5\mu} 
= -\e^uu^{\prime2}h_{5\mu} \, ,
\label{152747_23Nov16} \\
\hat{g}^{\alpha\beta}\Gamma^C_{\alpha5}\Gamma^D_{\beta\mu}h_{CD}
=&\hat{g}^{\alpha\beta}\Gamma^\gamma_{\alpha5}\Gamma^5_{\beta\mu}h_{\gamma5} 
=-\e^u\frac{u^{\prime2}}{4}h_{5\mu} \, , 
\label{152649_23Nov16} \\
\hat{g}^{\alpha\beta}\Gamma^C_{\alpha\mu}\Gamma^\delta_{\beta C}
h_{5\delta}=&\hat{g}^{\alpha\beta}\Gamma^\gamma_{\alpha\mu}
\Gamma^\delta_{\beta\gamma}h_{5\delta}
+\hat{g}^{\alpha\beta}\Gamma^5_{\alpha\mu}\Gamma^\delta_{\beta5}
h_{5\delta} \nn
=&\hat{g}^{\alpha\beta}\left[\hat{\Gamma}^\gamma_{\alpha\mu}
+\frac{1}{2}\left(\delta^\gamma_\alpha\partial_\mu u
+\delta^\gamma_\mu\partial_\alpha u
 -\hat{g}_{\alpha\mu}\hat{g}^{\gamma\lambda}\partial_\lambda u\right)\right]
\left[\hat{\Gamma}^\delta_{\beta\gamma}
+\frac{1}{2}\left(\delta^\delta_\beta\partial_\gamma u
+\delta^\delta_\gamma\partial_\beta u
 -\hat{g}_{\beta\gamma}\hat{g}^{\delta\lambda}\partial_\lambda u\right)
\right]h_{5\delta} \nn
&+\hat{g}^{\alpha\beta}\Gamma^5_{\alpha\mu}\Gamma^\delta_{\beta5}
h_{5\delta}  \nn
=&\hat{g}^{\alpha\beta}\hat{\Gamma}^\gamma_{\alpha\mu}
\hat{\Gamma}^\delta_{\beta\gamma}h_{5\delta}
+\frac{1}{2}\hat{\Gamma}^\gamma_{\alpha\mu}(\partial_\gamma u)
\hat{g}^{\alpha\delta}h_{5\delta}
+\frac{1}{2}\hat{\Gamma}^\delta_{\alpha\mu}
\hat{g}^{\alpha\beta}(\partial_\beta u)h_{5\delta}
 -\frac{1}{2}\hat{\Gamma}^\alpha_{\alpha\mu}(\partial_\lambda u)
\hat{g}^{\delta\lambda}h_{5\delta}  \nn
&+\frac{1}{2}\hat{g}^{\beta\gamma}\hat{\Gamma}^\delta_{\beta\gamma}
(\partial_\mu u)h_{5\delta}+\frac{1}{4}(\partial_\mu u)(\partial_\gamma u)
\hat{g}^{\gamma\delta}h_{5\delta}
+\frac{1}{4}(\partial_\mu u)(\partial_\beta u)\hat{g}^{\beta\delta}h_{5\delta}
 -(\partial_\mu u)(\partial_\lambda u)\hat{g}^{\delta\lambda}h_{5\delta}  \nn
&+\frac{1}{2}\hat{\Gamma}^\delta_{\beta\mu}\hat{g}^{\alpha\beta}
(\partial_\alpha u)h_{5\delta}
+\frac{1}{4}(\partial_\mu u)(\partial_\alpha u)\hat{g}^{\alpha\delta}h_{5\delta}
+\frac{1}{4}(\partial_\alpha u)\hat{g}^{\alpha\beta}(\partial_\beta u)h_{5\mu}
 -\frac{1}{4}(\partial_\mu u)(\partial_\lambda u)\hat{g}^{\delta\lambda}
h_{5\delta}  \nn
&-\frac{1}{2}\hat{\Gamma}^\delta_{\mu\gamma}\hat{g}^{\gamma\lambda}
(\partial_\lambda u)h_{5\delta}
 -\frac{1}{4}(\partial_\lambda u)\hat{g}^{\gamma\lambda}
(\partial_\gamma u)h_{5\mu}
 -\frac{1}{4}(\partial_\mu u)(\partial_\lambda u)\hat{g}^{\lambda\delta}
h_{5\delta}
+\frac{1}{4}(\partial_\mu u)(\partial_\lambda)\hat{g}^{\delta\lambda}h_{5\delta}  
\nn
&+\hat{g}^{\alpha\beta}\Gamma^5_{\alpha\mu}\Gamma^\delta_{\beta5}
h_{5\delta}  \nn
=&\hat{g}^{\alpha\beta}\hat{\Gamma}^\gamma_{\alpha\mu}
\hat{\Gamma}^\delta_{\beta\gamma}h_{5\delta}
+\frac{1}{2}\hat{\Gamma}^\gamma_{\alpha\mu}(\partial_\gamma u)
\hat{g}^{\alpha\delta}h_{5\delta}
+\frac{1}{2}\hat{\Gamma}^\delta_{\alpha\mu}\hat{g}^{\alpha\beta}
(\partial_\beta u)h_{5\delta}-\frac{1}{2}\hat{\Gamma}^\alpha_{\alpha\mu}
(\partial_\lambda u)\hat{g}^{\delta\lambda}h_{5\delta}  \nn
&+\frac{1}{2}\hat{g}^{\beta\gamma}\hat{\Gamma}^\delta_{\beta\gamma}
(\partial_\mu u)h_{5\delta}
 -\frac{1}{2}(\partial_\mu u)(\partial_\alpha u)\hat{g}^{\alpha\beta}h_{5\beta}
+\hat{g}^{\alpha\beta}\Gamma^5_{\alpha\mu}\Gamma^\delta_{\beta5}
h_{5\delta}  \nn
=&\hat{g}^{\alpha\beta}\hat{\Gamma}^\gamma_{\alpha\mu}
\hat{\Gamma}^\delta_{\beta\gamma}h_{5\delta}+\left(\frac{3\dot{a}\dot{u}}{a}
+\frac{\dot{u}^2}{2}\right)h_{50}\delta^0_\mu
 -\e^u\frac{u^{\prime2}}{4}h_{5\mu} \, , 
\label{152838_23Nov16} \\
\nabla^{(0)2}h_{5\mu}
=&\partial^2_5h_{5\mu}-\frac{u^{\prime\prime}}{2}h_{5\mu}
-u^\prime\partial_5h_{5\mu}+\frac{u^{\prime2}}{4}h_{5\mu} \nn
&+\e^{-u}\biggl[\hat{g}^{\alpha\beta}\partial_\alpha\partial_\beta h_{5\mu}
 -\hat{g}^{\alpha\beta}\partial_\alpha\hat{\Gamma}^\gamma_{\beta\mu}
h_{5\gamma}+\frac{\ddot{u}}{2}h_{5\mu} \nn
&-2\hat{g}^{\alpha\beta}\hat{\Gamma}^\gamma_{\beta\mu}
\partial_\alpha h_{5\gamma}-\dot{u}\hat{g}^{\alpha\gamma}
\partial_\alpha h_{5\gamma}\delta^0_\mu+\dot{u}\partial_0h_{5\mu}
 -\dot{u}\partial_\mu h_{50} \nn 
&-\hat{g}^{\alpha\beta}\hat{\Gamma}^\gamma_{\alpha\beta}
\partial_\gamma h_{5\mu}-\dot{u}\partial_0h_{5\mu}
+2\e^uu^\prime\partial_5h_{5\mu}  \nn
&+\hat{g}^{\alpha\beta}\hat{\Gamma}^\gamma_{\alpha\beta}
\hat{\Gamma}^\delta_{\gamma\mu}h_{5\delta}
 -\frac{\dot{a}\dot{u}}{a}h_{50}\delta^0_\mu+\frac{5\dot{a}\dot{u}}{2a}h_{5\mu}
+\frac{\dot{u}^2}{2}h_{5\mu}-\e^uu^{\prime2}h_{5\mu} 
 -\e^uu^{\prime2}h_{5\mu}  \nn
&-\e^u\frac{u^{\prime2}}{2}h_{5\mu}  
+\hat{g}^{\alpha\beta}\hat{\Gamma}^\gamma_{\alpha\mu}
\hat{\Gamma}^\delta_{\beta\gamma}h_{5\delta}+\left(\frac{3\dot{a}\dot{u}}{a}
+\frac{3\dot{u}^2}{2}\right)h_{50}\delta^0_\mu-\e^u\frac{u^{\prime2}}{4}
h_{5\mu}\biggr]  \nn
=&\partial^2_5h_{5\mu}-\frac{u^{\prime\prime}}{2}h_{5\mu}
+u^\prime\partial_5h_{5\mu}-\frac{5u^{\prime2}}{2}h_{5\mu}  \nn
&+\e^{-u}\biggl[\hat{g}^{\alpha\beta}\hat{\nabla}_\alpha\hat{\nabla}_\beta 
h_{5\mu}+\frac{\ddot{u}}{2}h_{5\mu}-\dot{u}\partial_\mu h_{50}
+\frac{5\dot{a}\dot{u}}{2a}h_{5\mu} \nn
&+\frac{\dot{u}^2}{2}h_{5\mu}+\frac{\dot{u}^2}{2}h_{50}\delta^0_\mu
+\frac{2\dot{a}\dot{u}}{a}h_{50}\delta^0_\mu-\dot{u}\hat{g}^{\alpha\gamma}
\partial_\alpha h_{5\gamma}\delta^0_\mu\biggr] \, .
\end{align}
Then by combining the above expressions, we find 
\begin{align}
&-R^{(0)}h_{5\mu}+R^{(0)55}h_{5\mu}+R^{(0)\nu}_\mu h_{5\nu}
 -2R^{(0)\ 5\nu5}_{\ \ \ \mu}h_{5\nu} \nn
=&\left[2u^{\prime\prime}+4u^{\prime2}
 -3\e^{-u}\left(\ddot{u}+\frac{\dot{u}^2}{2}+\frac{3\dot{a}\dot{u}}{a}
+\frac{2\ddot{a}}{a}+\frac{2\dot{a}^2}{a^2}\right)\right]h_{5\mu}  \nn
&+\left[-\frac{u^{\prime\prime}}{2}-u^{\prime2}+\e^{-u}\left(\frac{\ddot{u}}{2}
+\frac{\dot{u}^2}{2}+\frac{5\dot{a}\dot{u}}{2a}\right)\right]h_{5\mu}
+\e^{-u}\left(\ddot{u}-\frac{\dot{u}^2}{2}-\frac{\dot{a}\dot{u}}{a}\right)h_{50}
\delta^0_\mu+\e^{-u}\hat{R}^{\ \nu}_\mu h_{5\nu}  \nn
&+2\left(\frac{u^{\prime\prime}}{2}+\frac{u^{\prime2}}{4}\right)h_{5\mu} \nn
=&\left[\frac{5u^{\prime\prime}}{2}+\frac{7u^{\prime2}}{2}
 -\e^{-u}\left(\frac{5\ddot{u}}{2}+\dot{u}^2+\frac{13\dot{a}\dot{u}}{2a}
+\frac{6\ddot{a}}{a}+\frac{6\dot{a}^2}{a^2}\right)\right]h_{5\mu}
+\e^{-u}\left(\ddot{u}-\frac{\dot{u}^2}{2}-\frac{\dot{a}\dot{u}}{a}\right)h_{50}
\delta^0_\mu \, .
\label{170230_23Nov16} 
\end{align}

\end{document}